\documentclass[runningheads]{llncs}
\usepackage{graphicx}
\usepackage{marvosym}
\usepackage{booktabs}
\usepackage{multirow}
\usepackage{makecell}
\usepackage{hyperref}

\usepackage{bbding}
\begin{document}
\titlerunning{Automated Detection of Myopic Maculopathy in MMAC 2023}
\title{Automated Detection of Myopic Maculopathy in MMAC 2023: Achievements in Classification, Segmentation, and Spherical Equivalent Prediction}
%
%
\author{Yihao Li\inst{1,2}
\and
Philippe~Zhang\inst{1,2,3}
\and
Yubo~Tan\inst{4}
\and
Jing Zhang\inst{1,2}
\and
Zhihan Wang\inst{1,2}
\and
Weili~Jiang\inst{5}
\and
Pierre-Henri Conze\inst{1,6} 
\and
Mathieu Lamard\inst{1,2} 
\and
Gwenolé Quellec\inst{1} 
\and
Mostafa El Habib Daho \inst{1,2}\textsuperscript{(\Letter)}
}
\authorrunning{Y. Li et al.}
%
\institute{
LaTIM UMR 1101, Inserm, Brest, France \and
Univ Bretagne Occidentale, Brest, France \\
\email{mostafa.elhabibdaho@univ-brest.fr} \and
Evolucare Technologies, Villers-Bretonneux, France \and
University of Electronic Science and Technology of China, Chengdu, China \and
College of Computer Science, Sichuan University, Chengdu, China \and
IMT Atlantique, Brest, France \\
}
\maketitle              
\begin{abstract}
Myopic macular degeneration is the most common complication of myopia and the primary cause of vision loss in individuals with pathological myopia. Early detection and prompt treatment are crucial in preventing vision impairment due to myopic maculopathy. This was the focus of the Myopic Maculopathy Analysis Challenge (MMAC), in which we participated. In task 1, classification of myopic maculopathy, we employed the contrastive learning framework, specifically SimCLR, to enhance classification accuracy by effectively capturing enriched features from unlabeled data. This approach not only improved the intrinsic understanding of the data but also elevated the performance of our classification model. For Task 2 (segmentation of myopic maculopathy plus lesions), we have developed independent segmentation models tailored for different lesion segmentation tasks and implemented a test-time augmentation strategy to further enhance the model's performance. As for Task 3 (prediction of spherical equivalent), we have designed a deep regression model based on the data distribution of the dataset and employed an integration strategy to enhance the model's prediction accuracy. The results we obtained are promising and have allowed us to position ourselves in the Top 6 of the classification task, the Top 2 of the segmentation task, and the Top 1 of the prediction task. The code is available at \url{https://github.com/liyihao76/MMAC_LaTIM_Solution}.

\end{abstract}
\section{Introduction}

Myopia is a common eye disorder that affects millions of people worldwide~\cite{holden2016global}. It can develop into high myopia, leading to visual impairment, including blindness, due to the development of different types of myopic maculopathy~\cite{ikuno2017overview,silva2012myopic}. Myopic maculopathy is especially prevalent in countries such as Japan, China, Denmark, and the United States~\cite{ohno2015international,yokoi2018diagnosis}. The severity of myopic maculopathy can be classified into five categories~\cite{ohno2015international}: no macular lesions, tessellated fundus, diffuse chorioretinal atrophy, patchy chorioretinal atrophy, and macular atrophy. Three additional "Plus" lesions are also defined and added to these categories: Lacquer Cracks (LC), Choroidal Neovascularization (CNV), and Fuchs Spot (FS). 
Early detection and treatment are essential for preventing vision loss in people with myopic maculopathy. However, the diagnosis of myopic maculopathy is limited by the time-consuming and labor-intensive process of manually inspecting images individually. Therefore, developing an effective computer-aided system for diagnosing myopic maculopathy is a promising area of research. 

Deep learning (DL) methods have emerged as powerful tools in tackling challenges related to classification, segmentation, and prediction, demonstrating particular efficacy in medical imaging \cite{li2022segmentation,liu2022deepdrid,lahsaini2021deep}. In ophthalmology, DL has catalyzed advancements in diagnosing eye diseases, including Diabetic Retinopathy (DR) \cite{dai2021deep,quellec2021explain,daho2023improved} and myopia \cite{li2023predict_myopia,Zhang2023performances}. These recent studies have highlighted improvements in both the precision and efficiency of these diagnoses, underscoring the potential of DL in clinical applications.

Moreover, the field has seen innovative uses of abundant unlabeled data. Recognizing the cost and effort required for data labeling, researchers have pivoted towards self-supervised learning (SSL) approaches. These methods exploit unlabeled data for model pretraining in pretext tasks, subsequently applying them to target downstream tasks. This approach has proven effective in enhancing model performance and adaptability in various contexts \cite{zeghlache2022SSL,zeghlache:hal-04171357}.

Building upon these advancements in deep learning and the growing need for efficient, automated solutions in ophthalmology, the field has seen the inception of targeted initiatives like the Myopic Maculopathy Analysis Challenge (MMAC). The MMAC was organized to galvanize researchers worldwide to apply these innovative techniques in a focused setting. This challenge comprises three distinct tasks: (1) classification of myopic maculopathy, (2) segmentation of myopic maculopathy plus lesions, and (3) prediction of spherical equivalent, all utilizing a specially curated dataset of fundus images tailored to these tasks.

In alignment with these emerging trends and leveraging our expertise in deep learning, our team enthusiastically participated in all the tasks of this challenge. Our contributions were marked by notable achievements in each category:

\begin{itemize}
    \item In the classification task (\textbf{6th place}), we employed SimCLR, a contrastive learning method, that allowed the model to learn richer representations from the data. The integration of ensemble strategies, particularly when paired with SimCLR, further enhanced the model's robustness.
    \item In the segmentation task (\textbf{2nd place}), we designed and tested independent models for different lesion segmentation tasks. In addition, the Test Time Augmentation strategy we used boosted the performance of the models.
    \item In regression task (\textbf{1st place}), we focused on the analysis of the distribution characteristics of the dataset and designed the experimental protocol according to the distribution law of the dataset so that the deep regression model can learn and reason in a targeted manner. Furthermore, incorporating the model ensemble strategy increases the accuracy of the prediction.
\end{itemize}

\section{Materials and methods}

\subsection{Datasets}

The MMAC dataset is an extensive collection of color fundus images dedicated to research on myopic maculopathy. The dataset comprises fundus images gathered from various patients diagnosed with and without myopic maculopathy. \\
The sizes of each split of each task are summarized in Table \ref{Distribution}. To maintain the challenge's integrity and fairness, both the validation and test datasets are securely held and not released to participants. The validation and testing phases are executed on the organizer's side to ensure unbiased evaluation.\\

The first Task (Classification of Myopic Maculopathy) is focused on a five-category image classification. The categories are as follows:
\begin{itemize}
    \item Category 0: No macular lesions
    \item Category 1: Tessellated fundus
    \item Category 2: Diffuse chorioretinal atrophy
    \item Category 3: Patchy chorioretinal atrophy
    \item Category 4: Macular atrophy
\end{itemize}

The annotations were meticulously generated and reviewed manually by professional ophthalmologists. Two ophthalmologists annotated each image independently. In cases of discrepancies in labeling, a third senior ophthalmologist provided the final label. This rigorous process ensures the high quality and reliability of the dataset annotations.\\

The second Task (Segmentation of Myopic Maculopathy Plus Lesions) aims to segment three types of lesions:
\begin{itemize}
    \item Lacquer Cracks (LC)
    \item Choroidal Neovascularization (CNV)
    \item Fuchs Spot (FS)
\end{itemize}
An ophthalmologist first performed the lesion annotations. A second ophthalmologist refined these annotations in consultation with the first. Both these ophthalmologists have over five years of experience. A senior ophthalmologist with a decade of experience in ophthalmology reviewed and finalized the annotations.\\

For the third Task (Prediction of Spherical Equivalent), the true value of the spherical equivalent (SE) was ascertained using the corneal curvature computer refractometer TOPCON KR-8900. The SE was computed as: 
\begin{equation}
S E=S+\frac{1}{2} C
\end{equation}
where $S$  and $C$ are the spherical and cylinder diopter, respectively; these values were acquired through the computer refractometer.

\begin{table}[]
\centering
\caption{Distribution of color fundus images for different tasks.}
\label{Distribution}
\begin{tabular}{l|c|c|c}
\hline
\textbf{Task}                                                                                                                              & \textbf{Training Set} & \textbf{Validation Set} & \textbf{Test Set} \\ \hline
\textbf{Task1}- Classification of myopic maculopathy                                                                                      & 1143         & 248            & 915      \\ \hline
\begin{tabular}[c]{@{}l@{}}\textbf{Task2}- Segmentation of myopic maculopathy \\ plus lesions (Lacquer Cracks)\end{tabular}               & 63           & 12             & 46       \\ \hline
\begin{tabular}[c]{@{}l@{}}\textbf{Task2}- Segmentation of myopic maculopathy \\ plus lesions (Choroidal Neovascularization)\end{tabular} & 32           & 7              & 22       \\ \hline
\begin{tabular}[c]{@{}l@{}}\textbf{Task2}- Segmentation of myopic maculopathy \\ plus lesions (Fuchs Spot)\end{tabular}                 & 54           & 13             & 45       \\ \hline
\textbf{Task3}- Prediction of spherical equivalent                                                                                        & 992          & 205            & 806      \\ \hline
\end{tabular}
\end{table}

\subsection{Task 1: Classification of Myopic Maculopathy}

Several models were trialed for the classification task. These models included Resnet (18 and 50) \cite{He2015resnet}, ViT \cite{Dosovitskiy2022ViT}, and Swin \cite{Liu2021Swin}, among others. However, optimal results were achieved using a pipeline based on contrastive learning. This approach has recently gained traction for its ability to learn expressive representations from unlabeled data \cite{zeghlache2022SSL}\cite{zeghlache:hal-04171357}. Our implementation can be detailed as follows:

\subsubsection{Pretext Task: Contrastive Learning Framework (SimCLR)}

\begin{figure}[!htb]
\centering
\includegraphics[width=\textwidth]{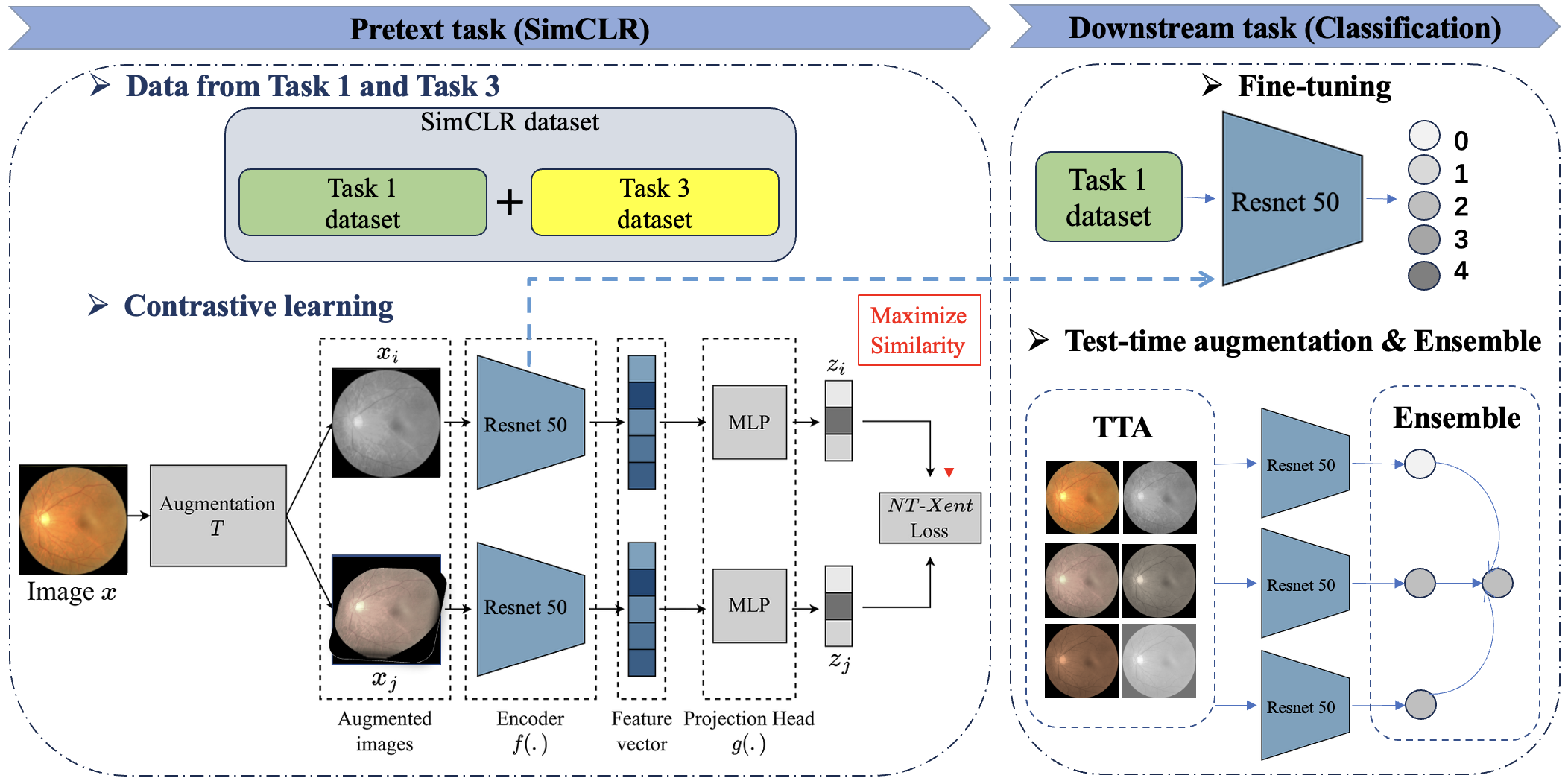}
\caption{Proposed pipeline for Task 1.} 
\label{pipeline_task1}
\end{figure}

As depicted in Fig.\ref{pipeline_task1}, our contrastive learning framework is rooted in the SimCLR architecture~\cite{Chen2022SimCLR}. The essence of SimCLR is to maximize the agreement between various augmented views of the same data instance through a contrastive loss in the latent space. A detailed breakdown of the utilized augmentations is provided in Table\ref{da3}. The architecture was supplemented with ResNet50\cite{He2015resnet} as its backbone, known for its depth and performance prowess in image-related tasks.

\begin{table}[!htb]
\centering
\caption{Data augmentations for Task 1 and Task 3.}\label{da3}
\resizebox{\linewidth}{!}{
\begin{tabular}{l|l|l}
\hline  
\textbf{Operator} & \textbf{Parameters} & \textbf{Probability}   \\
\hline
Flip & horizontal, vertical & 0.5\\
\hline
ShiftScaleRotate & shift\_limit=0.2, scale\_limit=0.1, rotate\_limit=45 & 0.5\\
\hline
RandomBrightnessContrast & brightness\_limit=0.2, contrast\_limit=0.2 & 1.0\\
\hline
RandomGamma & gamma\_limit=(80, 120) & 1.0 \\
\hline
CoarseDropout & \makecell[l]{max\_height=5, min\_height=1, max\_width=512, \\min\_width=51, max\_holes=5} &0.2\\
\hline
Sharpen &  alpha=(0.2, 0.5), lightness=(0.5, 1.0) & 1.0\\
\hline
Blur & blur\_limit=3 & 1.0 \\
\hline
Downscale &  scale\_min=0.7, scale\_max=0.9 & 1.0 \\
\hline

\end{tabular}}
\end{table}

\paragraph{Dataset Utilization:}
To make the most of the available data for the pretext unsupervised task, we amalgamated datasets from both Task 1 and Task 3. This approach not only broadened our data pool but also enabled the SimCLR model to capture a diverse range of features and representations.

\paragraph{Training Parameters:}
For training the SimCLR model, we standardized the image size to 256×256 pixels and utilized a batch size of 256. The temperature parameter was set at 0.07. Optimization was achieved using the AdamW optimizer coupled with the OneCycleLR scheduler. The learning rate was maintained at 0.001, and weight decay was set at 2.3e-05. The training persisted for a total of 2,000 epochs, ensuring ample time for convergence and representation learning.

\subsubsection{Downstream Task: Fine-tuning}

After the pretext task, we leveraged the learned representations for our primary objective, the classification of myopic maculopathy. For this supervised task, we fine-tuned the ResNet50 backbone extracted from the SimCLR architecture, excluding the projection head of SimCLR, which was discarded during this phase. The Task 1 dataset, accompanied by the provided labels, was employed for fine-tuning.

\paragraph{Checkpoint Strategy:}
To optimize our model's generalization, we employed a strategic checkpoint-saving approach. Checkpoints were saved based on various performance metrics, including Quadratic Weighted Kappa, Macro F1, and Macro Specificity. Additionally, a checkpoint capturing the average performance across these metrics was preserved. This strategy facilitated the eventual ensemble method, allowing for a harmonized prediction rooted in diverse evaluation criteria.

\paragraph{Test Stage:}
During the testing phase, we leaned into Test Time Augmentation (TTA) to enhance the robustness of our predictions. TTA has been shown to improve the generalization of models on unseen data. In conjunction with TTA, we employed an ensemble method drawing predictions from all saved checkpoints. This approach not only diversified our prediction source but also increased the reliability and accuracy of the final results.

\subsection{Task 2: Segmentation of myopic maculopathy plus lesions}

Segmentation of myopic maculopathy plus lesions in MMAC2023 is intended to detect pixel-level lesions, including LC, CNV, and FS. Table~\ref{Distribution} illustrates that the datasets required for different lesion segmentation tasks are different. As a result, it is difficult to obtain a unified segmentation model through multi-task learning in order to segment different lesions. Three independent segmentation models were used for this purpose. For the purpose of achieving optimal lesion segmentation, a data augmentation strategy has been proposed for the training of models on small datasets while backbone selection is performed. The TTA strategy was also incorporated to enhance the model's robustness.

\subsubsection{Data Split \& Augmentation} 

As with the Diabetic Retinopathy Analysis Challenge (DRAC) \cite{qian2023drac}, the segmentation tasks focus on the segmentation of lesions at the pixel level of 2D images, and there are a limited number of patients in the dataset. Following the best segmentation implementation \cite{kwon2022bag} in the DRAC challenge, we used all of the training data from the challenge to train our model. The data augmentation strategy outlined in Table~\ref{da2} was employed to avoid overfitting problems. The model will not encounter any original training samples due to the use of geometric transforms and pixel-wise transformations, which generate diverse input representations \cite{kwon2022bag}.

\begin{table}[htb]
\centering
\caption{Data augmentations for Task 2.}\label{da2}
\resizebox{\linewidth}{!}{
\begin{tabular}{l|l|l}
\hline  
\textbf{Operator} & \textbf{Parameters} & \textbf{Probability}   \\
\hline
Flip & horizontal, vertical & 0.5\\
\hline
ShiftScaleRotate & shift\_limit=0.2, scale\_limit=0.1, rotate\_limit=90 & 0.5\\
\hline
RandomBrightnessContrast & brightness\_limit=0.2, contrast\_limit=0.2 & 1.0\\
\hline
RandomGamma & gamma\_limit=(80, 120) & 1.0 \\
\hline
Sharpen &  alpha=(0.2, 0.5), lightness=(0.5, 1.0) & 1.0\\
\hline
Blur & blur\_limit=3 & 1.0 \\
\hline
Downscale &  scale\_min=0.7, scale\_max=0.9 & 1.0 \\
\hline
GridDistortion & num steps=5, distort limit=0.3 & 0.2 \\
\hline
CoarseDropout & \makecell[l]{max\_height=128, min\_height=32, max\_width=128, \\min\_width=32, max\_holes=3} &0.2\\
\hline
\end{tabular}}
\end{table}

\subsubsection{Backbone Selection} 

During the validation phase of the challenge, we extensively tested different segmentation backbones in order to determine the optimal one. Testing was conducted using Unet++ \cite{zhou2018unet++}, MAnet \cite{fan2020ma}, Linknet \cite{chaurasia2017linknet}, FPN \cite{kirillov2017unified}, PSPNet \cite{zhao2017pyramid}, DeepLabV3+ \cite{chen2018encoder} and $U^{2}$-Net \cite{qin2020u2} with the encoder of ResNet \cite{he2016deep} or EfficientNet \cite{tan2019efficientnet} architectures.

\subsubsection{TTA} The robustness of the model was improved by using TTA during inference. Each color fundus image was rotated by 90°, 180°, and 270° and then used in conjunction with the original image as input to the model. The final inference image was obtained by averaging the four different inference results after their restoration.

\subsection{Task 3: Prediction of spherical equivalent}

\begin{figure}[!htb]
\centering
\includegraphics[width=\textwidth]{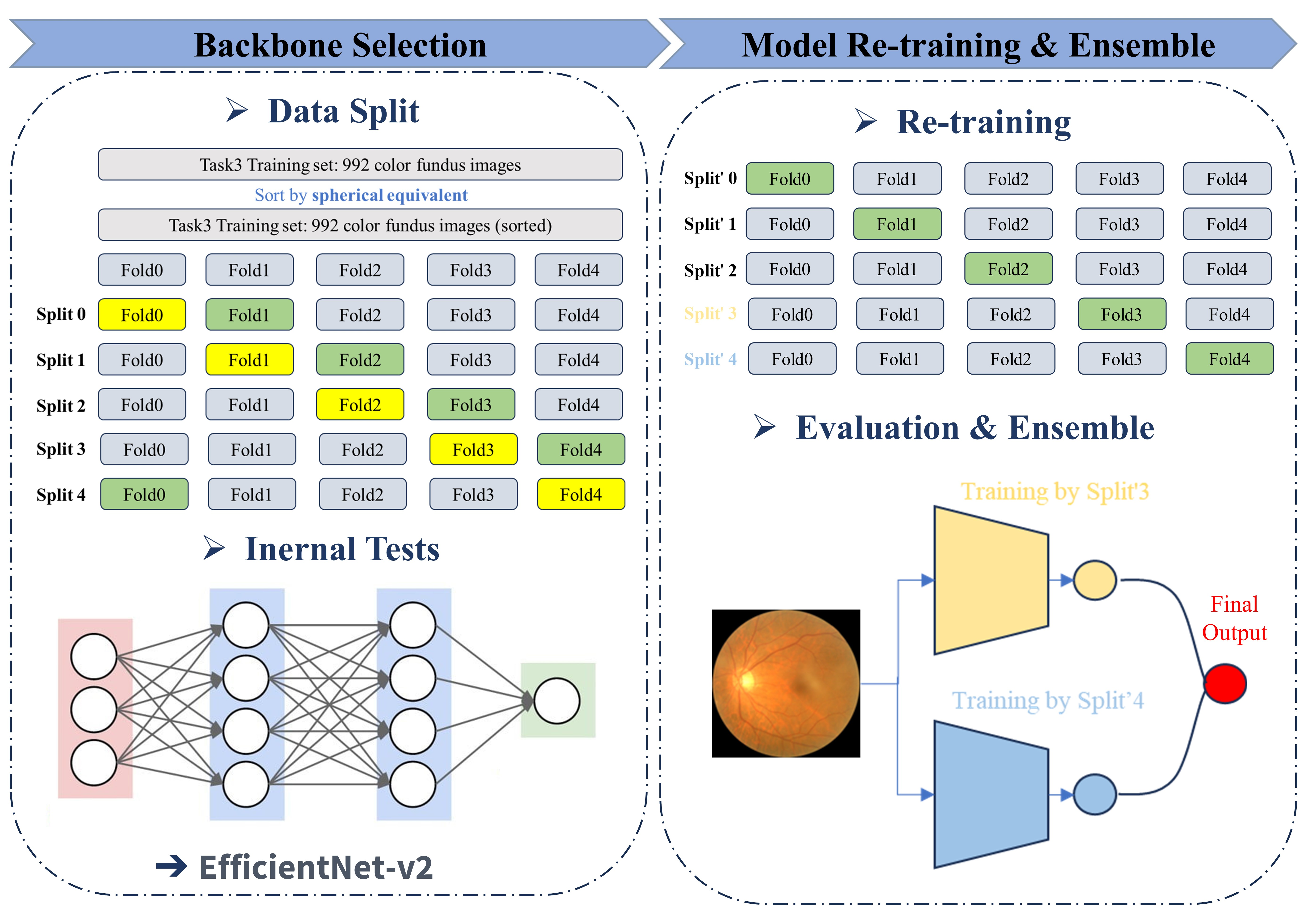}
\caption{Proposed workflow for Task 3. Gray folds represent the training set (internal), the validation set (internal) by green folds, and the test set (internal) by yellow folds.} \label{workflow3}
\end{figure}

The prediction of spherical equivalent can assist in diagnosing the risk of myopic maculopathy associated with increased degrees of myopia \cite{yokoi2018diagnosis}. Considering that there is a limited number of images in the training set and a limited number of submissions in the validation phase, we proposed the workflow shown in Figure~\ref{workflow3}. The following steps were tested for the prediction of spherical equivalent:

\begin{enumerate}

\item[(1)] Backbone Selection \\

To determine which backbone is most effective for the prediction task, we used five-fold cross-validation on the training set to assess various backbones' overall performance. In order to ensure a balanced data distribution, we sorted the data according to spherical equivalent values from smallest to largest. The training set of the challenge was split into three sets using five-fold cross-validation: an internal training set (3 folds), an internal validation set (1 fold), and an internal test set (1 fold).\\

Testing was performed on VGG \cite{simonyan2014very}, ResNet \cite{he2016deep}, DenseNet \cite{huang2017densely}, EfficientNet \cite{tan2019efficientnet}, and EfficientNetV2 \cite{tan2021efficientnetv2} architectures. The internal training set was used for training the model, the checkpoint was selected based on the R-Square value of the internal validation set, and the mean of the R-Square on the internal test of Split(0-4) was calculated to represent its overall performance. The list of data augmentation strategies is described in Table~\ref{da3}. \\

\item[(2)] Model Re-training \& Ensemble \\

In order to maximize the use of the training set data, we resplit the training set after the backbone testing. In this step, we used one fold of the internal test set from the previous step as the internal validation set and the remaining four folds as the new internal training set. The backbone that performs well is trained using the internal training set, while the checkpoints selected from the internal validation set are used to test the results on the challenge validation set. We got different models by training with different splits, and we used model ensemble \cite{lakshminarayanan2017simple} to further boost the performance by averaging their outputs. The performance of TTA during inference was also tested.

\end{enumerate}

\subsection{Implementation~Details}

The challenge comprised three diverse tasks, each with a set of unique requirements and evaluation parameters. Due to the inherent differences in the nature of these tasks, specialized evaluation metrics were formulated and employed for each. For task 1, the adopted evaluation indicators are Quadratic-weighted Kappa (QWK), F1 score, and Specificity. For task 2, the Dice Similarity Coefficient (DSC) is used to indicate the degree of coincidence of lesion segmentation. Finally, for the third task, the coefficient of determination R-squared, and the Mean Absolute Error (MAE) are employed to evaluate label regression's degree of correlation and distance.\\

\begin{table}[htb]
\centering
\caption{Implementation~details used in experiments.}
\label{configuration}
\resizebox{\linewidth}{!}{
\begin{tabular}{c|c|c|c}
\hline  
\textbf{Implementation} & \textbf{Task1} & \textbf{Task2} & \textbf{Task3}   \\
\hline
\textbf{Preprocessing} & None & \multicolumn{2}{c}{Normalize(mean=(0,0,0), std=(1,1,1))} \\
\hline
\textbf{Input size} & 512 $\times$ 512 pixels & \multicolumn{2}{c}{800 $\times$ 800 pixels} \\
\hline
\textbf{Backbone} & ResNet50 & \makecell[c]{MAnet\\(Encoder: ResNet34)} & \makecell[c]{EfficientNet-v2 \\(tf\_EfficientNetv2\_l)} \\
\hline
\textbf{Library} & timm & SMP & timm \\
\hline
\textbf{Pretrained weights} & Pretext Task & Imagenet & Imagenet\\
\hline
\textbf{Loss} &CrossEntropyLoss & DiceLoss+CrossEntropyLoss &  SmoothL1Loss \\
\hline
\textbf{Optimizer} &  \multicolumn{3}{c}{AdamW} \\
\hline
\textbf{Learning rate} & 1e-3 (OneCycleLR scheduler) & 1e-4 (w/o scheduler) & 2e-4 (w/o scheduler) \\
\hline
\textbf{Weight decay} & 2.3e-05 & 1e-2& 1e-2 \\
\hline
\textbf{Augmentation} & see Tab.~\ref{da3} & see Tab.~\ref{da2} & see Tab.~\ref{da3} \\
\hline
\textbf{Batch size} & 8 & 5 & 6 \\
\hline
\textbf{Epochs} & 200 & 1000 & 800 \\
\hline
\textbf{Train/Val split} & 0.8:0.2 & 1:0& 0.8:0.2\\
\hline
\textbf{Metric} & QWK, Macro F1, Macro Specificity &Dice &  R-squared \\
\hline
\end{tabular}}
\end{table}

In terms of computational infrastructure, the models and algorithms were implemented on a high-performance machine boasting 196 GB of RAM. The Nvidia GPU Tesla V100s with 32GB memory and NVIDIA A6000 with 48GB memory were employed to facilitate the intense calculations demanded by deep learning tasks. The software ecosystem was primarily built around PyTorch, a leading deep learning framework. Additional libraries like Timm (known for its efficient training routines and pre-trained models), Segmentation Models PyTorch (SMP) for the segmentation task, and Lightly for contrastive learning were incorporated to provide a robust and efficient system for the challenge's requirements.\\

Table~\ref{configuration} provides a brief description of our operators and detailed parameters for training. Unless otherwise specified, all experiments are conducted using reported configurations and parameters.

\section{Results}

\subsection{Task 1: Classification of myopic maculopathy}

Among the backbones we tested for classifying myopic maculopathy, ResNet50 delivered the most promising results, as presented in Table\ref{RES_test_1}. This superior performance prompted us to delve deeper into optimizing ResNet50 further. During this optimization, we integrated SimCLR as a Pretext task. The added value of SimCLR to our pipeline was evident, as demonstrated by the improved results compared to Resnet50 without a pretext task. This suggests that SimCLR effectively captures enhanced representations from unlabeled data, thus enriching our model's features and subsequently elevating its performance. Following this, the model was fine-tuned on the Task 1 dataset. \\
Further extending the model's capabilities, we experimented with several ensemble strategies:
\begin{itemize}
    \item Mean: The strategy involved selecting the classifier that showcased the best mean performance across the three metrics based on the validation set during training.
    \item All: This method calculated the mean of the logits outputs of each model before the Argmax operation, capturing a holistic insight from all models.
    \item Majority: A majority voting approach, this strategy collated predictions based on the predominant class predicted by all classifiers. 
\end{itemize}

Upon integrating ensemble strategies with ResNet50 optimized using SimCLR, we observed significant improvements in performance. The 'Majority' ensemble strategy combined with SimCLR achieved the highest Macro\_F1 score of 0.8176 and an overall score of 0.8881. Interestingly, the 'All' strategy demonstrated the peak QWK (0.9080) and Specificity (0.9427), emphasizing its capacity to capture comprehensive insights from the models. The ensemble method 'Mean' also displayed commendable results, with an overall score reaching 0.8781 when combined with SimCLR. It's evident from the data that the addition of ensemble strategies, particularly in tandem with SimCLR, significantly boosts the model's performance across various metrics.

\begin{table}[]
\centering
\caption{Task 1 validation results using different strategies}
\label{tab:my-table}
\begin{tabular}{c|c|c|c|c|c|c|c}
\hline
\textbf{Backbone}  & \textbf{Ensemble}  & \textbf{TTA} & \textbf{Pretext}  & \multicolumn{1}{c|}{\textbf{QWK}} & \multicolumn{1}{c|}{\textbf{Macro\_F1}} & \multicolumn{1}{c|}{\textbf{Specificity}} & \multicolumn{1}{c}{\textbf{score}} \\ \hline
Rexnet200 & \tiny{\XSolid}   &  \tiny{\XSolid}   & \tiny{\XSolid}          & 0.7524                   & 0.5873                  & 0.9126 & 0.7508      \\ \hline
Swin      & \tiny{\XSolid}   &  \tiny{\XSolid}   & \tiny{\XSolid}          & 0.8159                   & 0.6449                  & 0.9155  & 0.7921     \\ \hline
Resnet18  & \tiny{\XSolid}  & \tiny{\XSolid}   & \tiny{\XSolid}            & 0.8721                   & 0.6857                  & 0.9236  & 0.8271    \\ \hline
Resnet50  & \tiny{\XSolid} &  \tiny{\XSolid}   & \tiny{\XSolid}            & 0.8845                   & 0.7491                  & 0.9315  & 0.8550   \\ \hline
Resnet50  & Mean       &  \tiny{\XSolid}   & SimCLR             & 0.9030                    & 0.7926                  & 0.9388  & 0.8781    \\ \hline
Resnet50  & Majority    & \tiny{\XSolid}   & SimCLR      & 0.9067                   & \textbf{0.8176}         & 0.9400  & \textbf{0.8881}  \\ \hline
Resnet50  & All       &  \tiny{\XSolid}   & SimCLR              & \textbf{0.9080}         & 0.7954                  & \textbf{0.9427}  & 0.8821        \\ \hline
Resnet50  & Majority  & \tiny{\Checkmark}   & SimCLR       & 0.9028                   & 0.8049                  & 0.9385   & 0.8821   \\ \hline
\end{tabular}
\end{table}

As requested by the organizers, we submitted our four best-performing versions of our solutions for the final test, with these models being evaluated on a new, unseen dataset. Referring to Table\ref{RES_test_1}, we notice a slight decline in scores from validation to test phases, a common phenomenon due to the nuances of real-world data that may not be entirely captured in the validation set. The highest test scores were achieved using ResNet50 with the Majority ensemble method and TTA. This consistency from validation to test suggests that our methods are robust and not merely overfitting to the validation set.\\ 

\begin{table}[htb]
\centering
\caption{The performance of the different solution versions in Task 1 during the validation and testing phases}\label{RES_test_1}
\resizebox{\linewidth}{!}{
\begin{tabular}{c|c|c|c|c|c|c|c|c|c}
\hline  
\textbf{Phase}& \textbf{Ver.} & \textbf{Backbone} & \textbf{Ensemble} & \textbf{TTA}  & \textbf{Pretext} & \textbf{QWK}& \textbf{Macro\_F1} & \textbf{Specificity}& \textbf{score} \\
\hline
\multirow{4}*{Validation} &(1)& Resnet50 &Majority  &\tiny{\XSolid} & SimCLR& 0.9067                   & \textbf{0.8176}         & 0.9400  & \textbf{0.8881}\\
~ & (2) & Resnet50& Majority & \tiny{\Checkmark} & SimCLR& 0.9028                   & 0.8049                  & 0.9385   & 0.8821 \\
~ & (3) & Resnet50& All & \tiny{\XSolid} & SimCLR& \textbf{0.9080}         & 0.7954                  & \textbf{0.9427}  & 0.8821 \\
~ & (4) & Resnet50& Mean & \tiny{\XSolid} & SimCLR& 0.9030                    & 0.7926                  & 0.9388  & 0.8781    \\ \hline
\hline 
\multirow{4}*{Test} &(1)& Resnet50 &Majority  &\tiny{\XSolid} & SimCLR& 0.8811 & 0.7071 & 0.9373 & 0.8419\\
~ & (2) & Resnet50& Majority & \tiny{\Checkmark} & SimCLR& \textbf{0.8858} & \textbf{0.7081 }& 0.9396 &  \textbf{0.8445 }\\
~ & (3) & Resnet50& All & \tiny{\XSolid} & SimCLR& 0.8856 & 0.7044 &\textbf{ 0.9409 }& 0.8437 \\
~ & (4) & Resnet50& Mean & \tiny{\XSolid} & SimCLR& 0.8677 & 0.6942 & 0.9370 & 0.8330 \\
\hline
\end{tabular}}
\end{table}

\subsection{Task 2: Segmentation of myopic maculopathy plus lesions}

\begin{table}[!htb]
\centering
\caption{Results of Task 2 backbone selection on the validation set.}\label{bs2}
\resizebox{0.8\linewidth}{!}{
\begin{tabular}{c|c|c|c|c|c}
\hline  
\textbf{Backbone} & \textbf{Encoder} & \textbf{LC DSC} & \textbf{CNV DSC} & \textbf{FS DSC} & \textbf{Avg DSC} \\
\hline
UNet++ & EfficientNet-b0 & 0.7030	&0.5458&	0.7741	&0.6743 \\
\hline
UNet++ & EfficientNet-b1 & 0.6748&	0.5913&	0.7866&	0.6842\\
\hline
UNet++ & EfficientNet-b2 & 0.7081 & 0.5990 & 0.7881 & 0.6984 \\
\hline
UNet++ &  EfficientNet-b3  & 0.7158 & 	0.5516 &	0.7393 & 	0.6689\\
\hline
UNet++ & EfficientNet-b4 & 0.7087 & 0.6257 & 0.7123 & 	0.6823 \\
\hline
UNet++ & EfficientNet-b5 & 0.7051 &	0.5890 &	0.7956 &	0.6966\\
\hline
UNet++ & EfficientNet-b6 & 0.7203 &	0.5148 &	0.7940&	0.6764\\
\hline
UNet++ & EfficientNet-b7 & 0.6829 &	0.5895 & 0.8068 &	0.6931 \\
\hline
UNet++ & ResNet34 & 0.7303&	0.6339 &	0.8167 &	0.7270\\
\hline
UNet++ & ResNet50 & 0.7216&	0.4165&	0.8068&	0.6483 \\
\hline
UNet++ & ResNet101 & 0.7046&	0.6064&	0.7685&	0.6932\\
\hline
UNet++ & ResNet152 & 0.7055 &	0.5331 &	0.8306& 	0.6897\\
\hline
DeepLabV3+ & ResNet34 & 0.6986	&0.6351&	0.8347	&0.7228 \\
\hline
FPN & ResNet34 & 0.6969 &	0.5891 &	0.7769 &	0.6877 \\
\hline
Linknet & ResNet34 & 0.7304 &	0.5385&	0.7974 &	0.6888\\
\hline
PSPNet & ResNet34 & 0.7065 &	0.3974&	0.7908&	0.6316\\
\hline
MAnet & ResNet34 & 0.7573 &	\textbf{0.6885} &	\textbf{0.8498}&	\textbf{0.7652}\\
\hline
$U^{2}$-Net & - & \textbf{0.7651} &	0.3877&	0.7717&	0.6415 \\
\hline
\end{tabular}}
\end{table}

During the validation phase of the MMAC Challenge, we tested the performance of different backbones and encoders on the validation set, as shown in Table~\ref{bs2}. Based on the UNet++ structure, we compared the performance of ResNet and EfficientNet encoders. The ResNet34 encoder performed the best among them, and we used it to test the performance of other architectures. The segmentation of LC lesions using $U^{2}$-Net was found to be the most accurate, followed by segmentation using MAnet. MAnet demonstrated the best performance for segmenting both CNV lesions and FS lesions.\\

\begin{table}[htb]
\centering
\caption{The performance of the different solution versions in Task 2 during the validation and testing phases.}\label{eva2}
\resizebox{\linewidth}{!}{
\begin{tabular}{c|c|c|c|c|c|c|c|c|c}
\hline  
\textbf{Phase}& \textbf{Ver.} & \textbf{Model LC} & \textbf{Model CNV} & \textbf{Model FS} &\textbf{TTA} & \textbf{LC DSC} & \textbf{CNV DSC} & \textbf{FS DSC} & \textbf{Avg DSC} \\
\hline
\multirow{4}*{Validation} &(1)& $U^{2}$-Net & MAnet & MAnet & \XSolid & \textbf{0.7651} & \textbf{0.6885} & \textbf{0.8498} & \textbf{0.7678}\\
~ &(2)& $U^{2}$-Net & MAnet & MAnet & \Checkmark & 0.7367 & 0.6563 & 0.8024 & 0.7318\\
~ &(3)& MAnet & MAnet & MAnet & \XSolid &0.7573 & 0.6885 & 0.8498 & 0.7652 \\
~ &(4)& MAnet & MAnet & MAnet & \Checkmark &0.7563 & 0.6563 & 0.8024 & 0.7383 \\
\hline
\hline 
\multirow{4}*{Test} &(1)& $U^{2}$-Net & MAnet & MAnet & \XSolid & 0.6403 & 0.6250 & 0.8215 & 0.6956 \\
~ &(2)& $U^{2}$-Net & MAnet & MAnet & \Checkmark &  0.6682& 0.6557 & 0.8348 & 0.7196 \\
~ &(3)& MAnet & MAnet & MAnet & \XSolid & 0.6658 & 0.6250 & 0.8215 & 0.7041 \\
~ &(4)& MAnet & MAnet & MAnet & \Checkmark & \textbf{0.6838} & \textbf{0.6557} & \textbf{0.8348} & \textbf{0.7248} \\
\hline
\end{tabular}}
\end{table}

\begin{figure}[!htb]
\centering
\includegraphics[width=\textwidth]{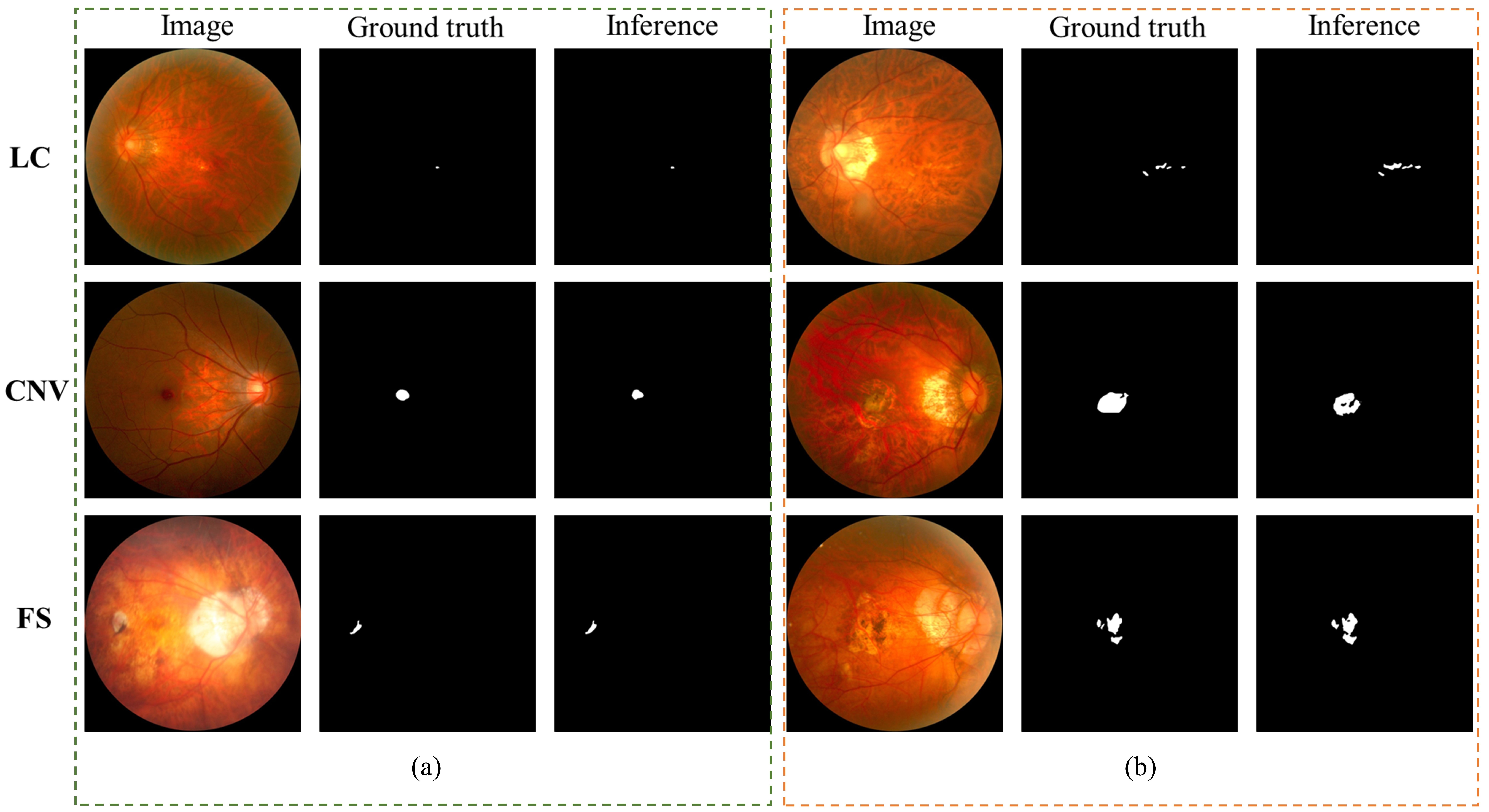}
\caption{Segmentation performance of MAnet on the validation set of Task 2.} 
\label{MAnet}
\end{figure}

Based on the results of the backbone tests, we selected the four best-performing versions as the solutions submitted in the testing phase of the challenge, as shown in Table~\ref{eva2}. The Ver. (1) combines the excellent performance of $U^{2}$-Net and MAnet on different segmentation tasks, thus performing well during the validation phase. Unfortunately, $U^{2}$-Net suffers from the overfitting problem due to the limited number of patients in datasets and therefore, performs poorly in the testing phase. Additionally, the tests indicate that the TTA approach significantly improves the robustness of the model and boosts the segmentation performance of the different submission versions in the testing phase. As a result, we found that the MAnet-based model with TTA strategy submission (Ver. (4)) performed optimally, with dice of 0.6838 for LC lesions segmentation, 0.6557 for CNV lesions segmentation, and 0.8348 for FS lesions segmentation on the test set. Figure~\ref{MAnet} illustrates the performance of our MAnet-based model on different lesion segmentation tasks. Our model demonstrates proficiency in segmenting smaller lesions, as depicted in Figure~\ref{MAnet}(a), but it shows limitations in accurately segmenting larger or more complex lesions, as observed in Figure~\ref{MAnet}(b).

\subsection{Task 3: Prediction of spherical equivalent}

\begin{table}[h]
\centering
\caption{R-Squared results of Task 3 with backbone selection on the internal test set.}\label{bs3}
\resizebox{0.8\linewidth}{!}{
\begin{tabular}{l|c|c|c|c|c|c}
\hline  
\textbf{Backbone(Timm)} & \textbf{Split0} & \textbf{Split1} & \textbf{Split2} & \textbf{Split3} & \textbf{Split4}& \textbf{Avg.} \\
\hline
vgg11 & 0.6773	& 0.7179	& 0.6808	& 0.6663	& 0.6241	&0.6733\\
\hline 
vgg16 & 0.5915 &	0.6507	&0.6821	&0.7699	&0.6753&	0.6739\\
\hline
resnet50 & 0.7061	& 0.7469	& 0.7328& 	0.6976	&0.6843	&0.7135\\
\hline
resnet152 & 0.7079	&0.6972	&0.6768	&0.7112	&0.7048	&0.6996\\
\hline
resnet200d & 0.7761	& 0.7675&0.7300 &	0.7494	& 0.7274	&0.7501\\
\hline
densenet121 & 0.7077 &	0.7289&	0.6821&	0.7303&	0.7057&	0.7109\\
\hline
densenet161 & 0.7543&	0.6865&	0.7094&	0.7449&	0.7337&	0.7258\\
\hline
densenet169 & 0.7416	&0.7367&	0.7166&	0.7285	&0.7410&	0.7329\\
\hline
densenet201 & 0.7405	&0.7347	&0.7094	&0.7172	&0.762	&0.7328\\
\hline
efficientnet\_b0 &  0.7318	& 0.7446	&0.7129&	0.7384&	0.7472&	0.7350\\
\hline
efficientnet\_b1 & 0.7108 &	0.7136	& 0.7084&	0.7098&	0.7535&	0.7192\\
\hline
efficientnet\_b2 & 0.7282 &	0.6952&	0.7027&	0.7554&	0.7486&	0.7260\\
\hline
tf\_efficientnet\_b6  & 0.7443 &	0.7918&	0.7098	&0.8264	&0.7272	&0.7599\\
\hline
tf\_efficientnet\_b7 & 0.7728	&0.7954	&0.7453	&0.7868	&0.7787	&0.7758\\
\hline
tf\_efficientnet\_b8 & 0.7980 &	0.8043	&0.8170	&0.8581&	0.7686	&0.8092\\
\hline
tf\_efficientnetv2\_s & 0.7765	&0.7746	&0.7206	&0.8069	&0.7520	&0.7661 \\
\hline
tf\_efficientnetv2\_l & 0.8147	&0.8374	&0.7801	&0.8226	&0.8336	&\textbf{0.8177}\\
\hline
tf\_efficientnetv2\_xl & 0.8115	&0.8354	&0.8005	&0.8006	&0.8166	&0.8129\\
\hline
\end{tabular}}
\end{table}

In order to evaluate the overall performance of different backbones on the internal test set, we first tested our proposed five-fold cross-validation method as shown in Table~\ref{bs3}. In light of the mean values of R-Squared across different Split internal test sets, we selected three backbones that provided a better performance: tf\_efficientnetv2\_l, tf\_efficientnetv2\_xl, and tf\_efficientnet\_b8. As the organizer's Python Packages currently do not support the current implementation of tf\_efficientnetv2\_xl, we chose to use tf\_efficientnetv2\_l and tf\_efficientnet\_b8 as backbones.\\

The dataset was resplit and then tf\_efficientnetv2\_l and tf\_efficientnet\_b8 were trained. As shown in Table~\ref{eva3}, some models that performed well on the new internal validation set were evaluated on the validation set of the competition. Based on the results, it can be seen that the tf\_efficientnet\_b8 model obtained by Split'1 training and the tf\_efficientnetv2\_l model obtained by Split'2 / Split'3 / Split'4 perform well on the validation set. These models are then ensembled. Our model ensemble methods performed well in both the validation and testing phases, improving prediction accuracy without overfitting. As a result of ensemble tf\_efficientnetv2\_l (Split' 3) and tf\_efficientnetv2\_l (Split' 4), the solution Ver. (3) obtained an R2 value of 0.8735 and an MAE value of 0.7080 on the test set.\\

\begin{table}[h]
\centering
\caption{The performance of the different solution versions in Task 3 during the validation and testing phases.}\label{eva3}
\resizebox{\linewidth}{!}{
\begin{tabular}{c|c|c|c|c|c|c}
\hline  
\textbf{Phase}& \textbf{Ver.} & \textbf{Ensemble} & \textbf{Model(Split)} &\textbf{TTA} & \textbf{R-Squared} & \textbf{MAE}\\
\hline
\multirow{9}*{\makecell[c]{ \\ \\ \\ \\Validation}} & - &  \tiny{\XSolid} & tf\_efficientnet\_b8(Split' 1) & \tiny{\XSolid} & 0.8230 & 0.6818 \\
\cline{2-7}
~ & - & \tiny{\XSolid} & tf\_efficientnetv2\_l(Split' 2) & \tiny{\XSolid} & 0.8526 & 0.6723 \\
\cline{2-7}
~ & (1) & \tiny{\XSolid} & tf\_efficientnetv2\_l(Split' 3) & \tiny{\XSolid} & 0.8622 & 0.6299 \\
\cline{2-7}
~ & - & \tiny{\XSolid} & tf\_efficientnetv2\_l(Split' 3) & \tiny{\Checkmark} & 0.8617 & 0.6307 \\
\cline{2-7}
~ & - & \tiny{\XSolid} & tf\_efficientnetv2\_l(Split' 4) & \tiny{\XSolid} & 0.8539 & 0.6570 \\
\cline{2-7}
~ & (2) & \tiny{\Checkmark} & \makecell[c]{tf\_efficientnet\_b8(Split' 1)+\\tf\_efficientnetv2\_l(Split' 3)}& \tiny{\Checkmark} & 0.8669& 0.6254\\
\cline{2-7}
~ & (3) &  \tiny{\Checkmark} & \makecell[c]{tf\_efficientnetv2\_l(Split' 3)+\\tf\_efficientnetv2\_l(Split' 4)}& \tiny{\XSolid} & 0.8734 & \textbf{0.6073} \\
\cline{2-7}
~ & - &  \tiny{\Checkmark} & \makecell[c]{tf\_efficientnetv2\_l(Split' 3)+\\tf\_efficientnetv2\_l(Split' 4)}& \tiny{\Checkmark} & 0.8705 & 0.6109 \\
\cline{2-7}
~ & (4) & \tiny{\Checkmark} & \makecell[c]{tf\_efficientnetv2\_l(Split' 2)+\\tf\_efficientnetv2\_l(Split' 3)+\\tf\_efficientnetv2\_l(Split' 4)}& \tiny{\XSolid} & \textbf{0.8745}& 0.6075\\
\hline
\hline 
\multirow{4}*{\makecell[c]{ \\ \\  \\Test}} &(1)& \tiny{\XSolid} & tf\_efficientnetv2\_l(Split' 3) & \tiny{\XSolid} & 0.8507 & 0.7627 \\
\cline{2-7}
~ & (2) & \tiny{\Checkmark} & \makecell[c]{tf\_efficientnet\_b8(Split' 1)+\\tf\_efficientnetv2\_l(Split' 3)}& \tiny{\Checkmark} & 0.8714 & 0.7258 \\
\cline{2-7}
~ & (3) &  \tiny{\Checkmark} & \makecell[c]{tf\_efficientnetv2\_l(Split' 3)+\\tf\_efficientnetv2\_l(Split' 4)}& \tiny{\XSolid} & \textbf{0.8735} & 0.7080 \\
\cline{2-7}
~ & (4) & \tiny{\Checkmark} & \makecell[c]{tf\_efficientnetv2\_l(Split' 2)+\\tf\_efficientnetv2\_l(Split' 3)+\\tf\_efficientnetv2\_l(Split' 4)}& \tiny{\XSolid} & 0.8732 & \textbf{0.7041} \\
\hline
\end{tabular}}
\end{table}

\section{Discussion and conclusions}

In this work, we presented our solutions for three tasks in the MMAC Challenge. For Task 1, ResNet50 emerged as the backbone of choice for classification. Its performance was substantially amplified upon incorporating SimCLR, a pretext task that effectively harnessed unlabeled data to enrich model representations. A deeper exploration of ensemble strategies, particularly the 'Majority' and 'All' methods, revealed significant performance boosts. These findings were validated on an unseen dataset, where our models demonstrated robustness, with the ResNet50 combined with the Majority ensemble method and TTA showcasing impressive consistency. For the final test rank in this task, our model secured the 8th position.\\

Segmentation of myopic maculopathy plus lesions (Task 2) provided its own set of challenges. The UNet++ structure fortified with ResNet34 encoder showcased promising results. However, a key revelation was the susceptibility of the $U^{2}$-Net model to overfitting, especially with limited datasets. Despite this setback, the MAnet-based model, augmented with the TTA strategy, emerged triumphant, achieving stellar dice scores for various lesion segmentations. This excellence in segmentation led our model to be ranked 2nd in the challenge.\\

The prediction of the spherical equivalent (Task 3) pivoted on our utilization of the tf\_efficientnetv2\_l and tf\_efficientnet\_b8 backbones, with the latter especially excelling in the validation phase. Embracing model ensembling further elevated our performance metrics. Notably, the ensemble of tf\_efficientnetv2\_l models derived from specific data splits yielded exceptional results on the test dataset, catapulting us to the top position for this task. \\

It should be noted that more models deserve further testing. As part of Task 3, we identified some backbone network architectures that performed well. However, we were unable to complete the tests due to limitations in the Python Packages provided by the organizer. Furthermore, nnUNet \cite{isensee2021nnu} is a common segmentation solution used in medical competitions. In light of the positive results achieved by nnUNet in the DRAC Challenge \cite{li2022segmentation}, it raises the possibility that nnUNet can also perform well in the MMAC Challenge.

\section*{Acknowledgements}
The work was conducted in the framework of the ANR RHU project Evired. This work benefited from state aid managed by the French National Research Agency under the ``Investissement d'Avenir'' program, reference~ANR-18-RHUS-0008.

\newpage

%
%
%
\bibliographystyle{splncs04}
\bibliography{mybibliography}

\end{document}